# Effects of low-level deuterium enrichment on bacterial growth

*Xueshu Xie[1] and Roman A. Zubarev[1]\**

1. Division of Physiological Chemistry I, Department of Medical Biochemistry and Biophysics, Karolinska Institutet, Stockholm, Sweden

\*Corresponding author: Roman.Zubarev@ki.se   phone/fax +46 8 524 87 594

**Abstract**

Using very precise (±0.05%) measurements of the growth parameters for bacteria *E. coli* grown on minimal media, we aimed to determine the lowest deuterium concentration at which the adverse effects that are prominent at higher enrichments start to become noticeable. Such a threshold was found at 0.5% D, a surprisingly high value, while the ultralow deuterium concentrations (≤0.25% D) showed signs of the opposite trend. Bacterial adaptation for 400 generations in isotopically different environment confirmed preference for ultralow (≤0.25% D) enrichment. This effect appears to be similar to those described in sporadic but multiple earlier reports. Possible explanations include hormesis and isotopic resonance phenomena, with the latter explanation being favored.

**Introduction**

Since the discovery of D$_2$0 (heavy water) in 1932 by Urey, Brickwede and Murphy [1], its biological effects have attracted a great deal of researchers' interest. Already early experiments have revealed that deuterium has profound effect on living organisms. Between 1934 and the beginning of the second World war in 1939, a total of 216 publications appeared dealing with biological effects of deuterium [2]. Excess of deuterium in water was found to cause reduction in synthesis of proteins and nucleic acids, disturbance in cell division mechanism, changes in enzymatic kinetic rates and cellular morphological changes [3,4]. Yet it is possible to grow microorganisms (e.g. some variants of *Escherichia coli*) in a highly substituted medium, and achieve almost complete deuterium substitution [5,6].

The biological effects of stable isotopes are usually observed shortly after the microorganism is placed in an isotopically different medium [7]. At first, prokaryotic cells experiences an "isotopic shock" manifested through the growth arrest and morphology changes. After a period of adaptation ("lag phase"), growth resumes, but the rate is usually slower than in normal isotopic environment [8]. The changes in the growth rate can be explained by the impact of isotopic substitutions on the kinetics of enzymes [9], pattern of hydrogen bonds and similar relatively subtle but cumulatively potentially important effects.

The Katz group who have studied multiple heavy-isotope ($^{13}$C, $^{15}$N and $^{18}$O) substitutions in *Chlorella vulgaris* grown in heavy water found that all additional isotopic substitutions resulted into abnormal effects in cell size, appearance, growth rate and division [10]. The effects were progressively stronger as the isotopic composition deviated from the normal. The authors concluded that "*organisms of* *different isotopic compositions are actually different organisms, to the degree that their isotopic compositions are removed from naturally occurring compositions*" [10].

The question is whether this conclusion made for heavy enrichments levels (>50%) remains true for low levels of enrichment (<10%). Since isotopic compositions of microorganisms grown in a slightly enriched environment showed deficit of heavy isotopes [11], one can reasonably assume that even low levels of deuterium enrichment may cause biological effects. Indeed, the earliest works on biological effect of deuterium reported in the 1930s were performed using low levels of deuterium, as high enrichment was still not available. Barnes *et al.* studied the physiological effect of low deuterium concentration on the growth of *Spirogyra*, flatworms and *Euglena*. They found an increased growth of these three organisms at 0.06% of D in water compared to ordinary water that contains 156 ppm D (0.0156%). More specifically, they observed increased cell division for *Euglena*, longer longevity for flatworms, and both less cell disjunction and greater longevity for *Spirogyra* [12–15]. Lockemann and Leunig's study on the effect of heavy water with less than 0.54% D upon *E. coli* and *Pseudomonasa* revealed that concentrations as low as 0.04% D favored growth [16]. Curry *et al.* observed ca. 10% faster growth for *Aspergillus* at 0.05% D, but the result was not statically significant due to large errors of measurements [17]. After the first half of 1930s, the research focus has shifted to the effects of highly enriched deuterium, which were more pronounced and largely negative. The interested to low deuterium enrichment has returned in the 1970s and 1980s, when Lobyshev *et al.* studied the Na, K-ATPase activity at different concentration of deuterium and found it to increase at low deuterium concentrations, with a maximum reached at 0.04-0.05% [18,19]. Lobyshev *et al.* then performed experiments with regeneration of hydrioid pohyps *Obelia geniculata* in a wide range of deuterium added to sea water and found faster regeneration at and below 0.1% D [20]. Somlyai *et al.* have shown that 0.06% D in tissue culture activated the growth of $L_{929}$ fibroblast cell lines [21]. Nikitin *et al.* have studied growth of bacteria with different membrane lipid composition in liquid media in a range of deuterium concentration in water varied from 0.01% to 90% and observed pronounced activation in growth for *Methylobacterium organopholium* and *Hyphomonas jannaschiane* at around 0.01% enrichment [22]. Recent studies on the impact of $D_2O$ on the life span of *Drosophila melanogaster* revealed the biggest positive effect of deuterium at the lowest D concentration tested in that work, 7.5% of enrichment [23].

In the current study, we probed enrichment levels starting from 0.03%, i.e. double the normal deuterium abundance of 156 ppm. Disregarding the earliest reports from 1930s on the biological effects of 0.06% deuterium in water which suffered from the lack of statistical analysis and proper controls, we expected the size of the biological effects to be commensurable with the enrichment levels, i.e. to be exquisitely small. To address the expected minute size of the phenomena, we developed a very sensitive method for detecting the biological effects of unspecific deuterium incorporation in the model organism. *E. coli* was chosen as such because of the ease of handling, robustness and speed of growth as well as the ability to thrive on minimal media with easily changeable isotopic composition.

The applied method utilizes robotized sample preparation, automated and massively parallel data acquisition (hundreds of individual experiments per concentration point) and measures three independent parameters per growth curve: maximum growth rate, the lag phase duration and maximum density (Figure 1). Massive parallel measurement approach is achieved using the BioScreen C automated fermentor that affords up to 200 experiments run at the same time. Every second experimental well in a 100-well plate was kept at standard isotopic conditions (Figure S1), and the data from each "test" well was normalized by that of the neighboring standard well, providing the accuracy of relative measurements close to those achieved with internal standard. Using massive statistics, we achieved the precision of relative measurements close to 0.05% (standard error).

**Materials and Methods**

*Chemicals and Materials*

Glycerol stock of *E. coli* BL 21 strain (stored at -80 $^o$C) was obtained from the microbiology lab of our department. The M9 minimal media [24] was prepared using D-glucose ($C_6H_{12}O_6$), disodium hydrogen phosphate ($Na_2HPO_4 \cdot 2H_2O$), monopotassium phosphate ($KH_2PO_4$), sodium chloride (NaCl), magnesium sulfate ($MgSO_4$), calcium chloride ($CaCl_2$), ammonia chloride ($NH_4Cl$), all purchased from Sigma-Aldrich (Schnelldorf, Germany), and distilled water prepared with a Milli-Q device from Millipore (Billerica, MA, USA). Heavy water (99.9% of $^2$H) was also purchased from Sigma-Aldrich (Schnelldorf, Germany).

Vacuum filtration system with 0.2 μm polyethersulfone (PES) membrane for bacteria media sterilization was purchased from VWR (Stockholm, Sweden). Petri dishes (90 x 14 mm) and inoculating sterile loops were purchased from Sigma-Aldrich. Sterile plastic conical tubes (50 mL and 15 mL) for sample preparation were purchased from Sarstedt (Nümbrecht, Germany). The BioScreen C automatic fermentor was obtained from Oy Growth Curves AB Ltd (Helsinki, Finland).

*M9 minimal media preparation*

5-time concentrated M9 minimal salts stock solution was prepared by dissolving 42.5 g $Na_2HPO_4 \cdot 2H_2O$, 15 g $KH_2PO_4$ and 2.5 g NaCl in Milli-Q water to a final volume of 1000 mL. The solution was then sterilized by autoclaving and stored at 4 $^o$C for further use. To prepare M9 minimal media, the salts stock solution was diluted five times in Milli-Q water.

M9 minimal media were prepared by mixing the following components: 800 mL Milli-Q water, 200 mL M9 concentrated salts stock solution, 2 mL of 1 M $MgSO_4$ solution, 0.1 mL of 1 M $CaCl_2$ solution, 5 g D-glucose ($C_6H_{12}O_6$) and 1 g $NH_4Cl$.

*Preparation of streak agar plates*

M9 minimal media agar plates were prepared by dissolving 3 g of agar powder in 200 mL M9 minimal media. The obtained mixture was sterilized by autoclaving, then cooled down to ca. 60 $^o$C and finally poured into Petri dishes (ca. 15 mL agar solution per plate). The agar plates were allowed to solidify at room temperature for ca. 10 min, sealed with parafilm and stored at 4 $^o$C till further use.

*E. coli* streak agar plates were prepared by streaking [25] *E. coli* from -80 $^o$C glycerol stock onto M9 minimal media agar plate followed by 40-hour incubation at 37 $^o$C to form visible isolated colonies. Streak agar plates were stored for experiments for maximum one week at 4 $^o$C.

*Measurement of E. coli growth*

From the *E. coli* agar plate, one isolated colony was picked with a sterile loop into 5 mL M9 minimal media and incubated at 37 $^o$C while shaking with 200 r.p.m for 5-6 hours until it reached its early exponential phase with optical density ($OD_{590}$) around 0.2, measured with Colorimeter WPA CO75 (York, UK).

Sample preparation workflow is shown in Figure 2. In each experiment, four stock solutions were used. Stock A for preparing sample $S_A$, and stock B for preparing sample $S_B$ were obtained by mixing M9 minimal media with sterilized heavy water at a certain ratio (Table 1). For the preparation of stock solutions of standard A and standard B, M9 minimal media were mixed with sterile Milli-Q water at the same ratio as stock A and stock B. The final solutions were dispensed into the honeycomb well plates using programmed robotic system (Tecan, Genesis RSP 150, Männedorf, Switzerland).

A 20 µL aliquot of the incubated *E. coli* culture (O.D. ≈ 0.2) was diluted in 45 mL M9 minimal media for robotic sample preparation. First, 300 µL M9 minimal media without bacteria were introduced into each of the border wells ("edge cells") on both plates $P_A$ and $P_B$ (72 wells in total) to serve as blank samples (no color code in Figure S1). Second, 30 µL (10 µL for samples with <4% enrichment of deuterium) aliquot of stock A was dispensed into each "sample" well (marked with purple on plate $P_A$ in Figure S1) to prepare 32 replicates of sample $S_A$. Third, 30 µL (10 µL for samples with <4% enrichment of deuterium) aliquot of stock solution of standard A was added into each "standard" well (marked with yellow on plate $P_A$ in Figure S1) to prepare 32 reference standards on plate. In the same way, wells were filled on plate $P_B$. Finally, 270 µL (290 µL for <4% D) of the diluted *E. coli* culture was dispensed into each well except blank wells. In total, 32 replicate pairs of "sample" and "standard" wells were prepared on each plate. Sample configuration is shown in Figure S1.

Bioscreen C monitors *E. coli* concentration in each well by measuring turbidity (with wide band filter 420–580 nm) in it at 39 °C with continuous bacterial agitation by shaking. In our experiments, turbidity was sampled every six minutes and was monitored for ca. 24 hours.

*E. coli adaptation to different deuterium content*

One *E. coli* colony was inoculated into 5 mL M9 minimal media (normal isotopic condition) from a fresh M9 minimal media agar plate, and cultured at 37 $^o$C until O.D. 1.0. For every deuterium concentration point (156 ppm, 0.03%, 0.25%, and 1%), 2 µL *E. coli* culture was diluted 2500 times into 5 mL M9 minimal media with the corresponding D content and cultured overnight at 37 $^o$C. After the culture reached its maximum density (O.D. ≈ 1.4), for each *E. coli* culture, 10 µL was then diluted 500 times into 5 mL media with the same D content, and culturing continued at 37 $^o$C. In this way, each *E.coli* culture was diluted 500 times twice a day, with nine generations grown between dilutions. The process continued until the 400$^{th}$ generation was obtained, after which the difference in the rate of growth of the adapted bacteria on the corresponding content of deuterium versus normal D content was measured using Bioscreen C.

*Data Analysis*

Using Microsoft Excel, the logarithm of turbidity was plotted against time (Figure 1). The slope for every 8-h interval was calculated, and the maximum value was determined. The extrapolation of the line with maximum slope to the background level of turbidity gave the lag time. The maximum turbidity for each replicate was taken as the maximum density. The obtained three values for each growth curve were treated in the same way as below.

For each "sample" A, the obtained value was normalized by that of the "standard" B. To minimize the influence of nonstatistical outliers that could arise due to gross errors in sample preparation and handling (e.g. robot or BioScreen C malfunction), the 32 replicates were divided into 4 groups according to their positions on the honey comb well plate (group 1: columns 1 and 2; …, group 4: columns 7 and 8). In each group, the median of the eight values was calculated and then the four medians were averaged to obtain the value for a given plate and its standard deviation.

Altogether, nine independent 32-replicate experiments were performed for each D content point and six experiments per each D point for adapted bacteria. To compensate for small systematic errors of measurement (e.g., differences in the geometry of the honeycomb wells, position-dependent sensitivity of the BioScreen C detector, etc.), the obtained average values of the maximum growth rate, lag time as well as the maximum density within each experiment were normalized to that of the sample with terrestrial D composition (156 ppm). The final result was obtained when the average of 9x4=36 (6x4=24 for adapted bacteria) median values, and the corresponding standard error, were calculated. Since after normalization, the values for terrestrial composition were all unities, the p-values for non-terrestrial compositions were calculated using two-tailed, paired Student's t-test comparing the relative value R and its inverse 1/R.

**Results and Discussion**

*High enrichment (50% D).* To learn the behavior of the growth parameters at high enrichment, bacteria were grown first at 50% D. The growth curve showed significant deviation from the normal (Figure 1), with the growth rate reduced by ca. 5%, maximum density increased by ca. 15% and the lag phase extended by almost 40% (Figure S2). While reduction in the growth rate and lag time extension were expected based on abundant previous research, the increase in maximum density was surprising: at adverse growth conditions, the maximum density usually follows the declining growth rate trend. The increase in maximum density at 50% D, or at least the magnitude of the increase, could be overestimated by the optical density measurements if the optical properties of the deuterated bacteria are different than that of bacteria with normal D content. Indeed, literature suggests that the size and shape of the bacteria may be quite distorted when grown in a highly deuterated media, especially in the beginning of growth [3,4,7]. If this is the case, and the maximum density at 50% D is overestimated, then the growth rate reduction at 50% D may be underestimated in our measurements. However, both these possible effects should be relatively small at <10% D, and especially at <1% D. Besides, our prime goal was to identify the lowest concentration of deuterium at which biological effects become detectable. Thus we did not make any corrections for these hypothetical effects in low-enrichment measurements.

*Low enrichment (<10% D).* Growth behavior of *E. coli* in minimal media under ten different content of deuterium (including 156 ppm, 0.03%, 0.06%, 0.1%, 0.25%, 0.5%, 1%, 2%, 4%, 8%) were investigated, starting with 156 ppm of deuterium (terrestrial condition) and increasing the content of deuterium by a factor of 2 until ca. 10% D is reached. In each run of the BioScreen C instrument, *E. coli* was grown in minimal media on two honey comb well plates each containing wells with two different contents of deuterium, one always being the normal D content (156 ppm), giving 32 replicates for each plate (*vide infra*). This way, five runs were required to cover the entire range of deuterium content, which constituted one experiment.

Figure 3A shows the normalized maximum growth rate. The curve is trending downwards, but only at the highest enrichment the statistical significance at $p<0.05$ is reached. The observed trend significantly (by a factor of 4) lags behind the linear extrapolation of the 50% D results to low-enrichment levels (i.e., a 5% slower growth at 50% D means growth rate $\approx 1 - 0.001*\%D$).

In contrast, the lag phase curve (Figure 3B) fluctuates at ultralow enrichment levels slightly below the normal level. But after 0.1% D, the lag time increases dramatically, reaching statistical significance at 0.5% D. The curve lags behind the linear prediction by a factor of two. Even if linear extrapolation is made based on 8% D results instead of 50% D, the actual trend remains below the predicted one.

The maximum density curve (Figure 3C) shows surprising behavior: at the lowest enrichment level of 0.03% D amounting to only twice the normal D content, there is a small but significant ($p<0.05$) peak corresponding to a $\approx 0.16\%$ higher density. The increase in the maximum density continues at higher D content, and vanishes at 0.25% D, to increase again above the significance level at 1% D. For the interval 0.2% - 4% D, the maximum density curves lags behind the linear extrapolation, but at 8% D it actually exceeds it, increasing by 3.2% compared to the predicted 2.4%.

Summarizing the above observations, the effects typical for high D enrichment (slower growth, longer lag and higher maximum density), start to be noticeable at 0.5% D. This is a surprisingly high threshold, given that the 0.05% precision of the measurements would allow us to detect with statistical significance as small changes in growth parameters as 0.1%. According to the 50% D results, if the effect of D content on *E. coli* growth was linear, the expected detection threshold for biological effects would be around 0.1-0.2% D. That no significant change was detected at 0.25% D indicates the presence of a previously unreported threshold for isotopic phenomena. Moreover, the results hint on the existence of an "inverted region" at ultralow enrichment levels with statistically significant fluctuations in the direction of higher comfort, rather than adverse reaction as at ≥0.5% D. The existence of such a region would be in agreement with previous studies, even though many of the actually found much bigger effects than were observed in our experiments [12–22].

*Validity of ultralow enrichment effect.* The biological effect at <0.25% D in our study does not result from one or two aberrantly high measurements: in six out of nine constituent experiments, the maximum density was enhanced compared to normal % D. Another way to test the validity is to investigate the variability of measurements: usually, stronger biological effect results in larger variability. Indeed, the standard error of the maximum density determination at normal D content was 0.05%, while for 8% D it was five times larger, 0.25%. At 0.03% D, the corresponding standard error was 0.08%, which is the highest value for all measurements below 1% D. Also for the other two growth parameters, the maximum growth rate and the lag phase, the standard error at 0.03% D was higher than at normal conditions. Curiously, of all the data points, only 0.03% D shows an increase (however slight and statistically insignificant) in the maximum growth rate on Figure 3A. Another supporting evidence in favor of the opposite trend at ultralow enrichment is the fact that, at such enrichment, all three growth parameters lag behind the trend extrapolated from 50% D measurements, and two out of three (lag phase and maximum density) – of the trend extrapolated from 8% D results. These results speak in favor of the validity of the biological effect at 0.03% D.

The effect at the ultralow enrichment levels, assuming it is real, in unlike that at high enrichments. While the growth rate does not change, the lag phase shortens (although insignificantly in statistical sense), and the maximum density increases above the expected trend. Overall, it's a signature of growth in a more comfortable environment. In contrast, at high enrichments the signature is definitely that of a more hostile environment, with a very long lag phase and slower growth. What is common for both regimes, though, is the increased maximum density, but the origins of this effect may differ (*vide supra*).

How to explain the presumed effect of ultralow deuterium enrichment on bacterial growth? The previous research has not produced any convincing explanation [12–22], which probably was one of the factors why this research has largely became forgotten. One of possible explanations is the effect of hormesis. The concept of hormesis suggests that, at low dosages, the effects of adverse agents (e.g. chemical toxins or radiation) can be stimulating for growth. Hormetic effect on *E. coli* growth has previously been detected at low concentrations of chemical toxins [26]. However, the size of the effect in our study (≈0.16% in maximum density) is disproportionally large compared to the enrichment degree (0.032% D, or +0.016% D compared to normal composition, Figure S3), which speaks against the role of hormesis in the studied situation.

*Adaptation to ultralow enrichment conditions*. Hormesis is usually explained by activation of protecting mechanisms that, in effect, boost the growth at low concentrations of adverse agents. But gradual adaptation of the organism to slightly adverse environment should increase the threshold for activation and thus reduce this effect. To test this hypothesis, we grew bacteria for ca. 400 generations at normal isotopic conditions, 0.03% D, 0.25% D and 1% D, changing the media every 12 hours (Figure S4). After that, the growth parameters in the same media were measured, using as control the same adapted bacteria growing at normal isotopic conditions. Adaptation has dramatically reduced the lag time from 10-11 h to 6-7 h, and increased the maximum growth rate as well as the maximum density by about 15-20% (Figure S5). The directions of the changes in growth parameters are consistent with bacteria feeling much more comfortable in the minimal media after adaptation.

In the isotopically different environment, the growth rate has increased by 0.10-0.15% at ultralow enrichment ($p<0.05$ at 0.25%, Figure 4A), while the lag time has increased insignificantly by the same amount (Figure 4B). The maximum density has shown most significant changes of ca. 0.2% ($p<0.05$, Figure 4C) at 0.25% D, but was elevated also at 0.03% D. Overall, two out of three parameters showed statistically significant changes characteristic for better adaptation in isotopically different environment compared to normal media. Note that at 1% D this effect disappeared.

Strictly speaking, deuterium is not toxic for *E. coli*, as this organism can grow at very high enrichment levels. Even at 50% D, the maximum growth rate is only slightly (5%) lower than at normal conditions. Therefore, deuterium enriched medium is more correctly characterized as an unusual, rather than toxic, environment. Thus the applicability of the hormetic explanation to ultralow deuterium enrichment is questionable.

An alternative to hermetic explanation could be provided by the isotopic resonance hypothesis [27]. Briefly, the hypothesis suggests that, at certain isotopic compositions of the elements C, H, N and O that together compose ≥96% of the living matter of prokaryotes and many eukaryotes, the rate of all chemical and biochemical reactions, including the bacterial growth, should increase. Normal terrestrial isotopic compositions are very close to one such resonance. The existing small deviation from the perfect resonance, suggests the hypothesis, can be corrected by "tuning" the isotopic composition of any of the elements C, H, N and O. After such tuning the growth rate of all organisms should slightly increase. The size of the effect is not predicted, but the required value of the D content for the perfect "terrestrial" isotopic resonance is close to 0.03% [28]. Since the isotopic composition of water inside the *E. coli* bacteria has much lower deuterium enrichment compared to the growth media [29], the resonance conditions should occur in the region 0.03-0.06% of D. Such an agreement between the predictions of the hypothesis and the experiment could be spurious, but it receives strong support from earlier studies [12–22]. Taken together, these results suggest that the growth behavior of *E. coli* and other organisms at ultralow enrichment levels well worth exploring further, using even more precise measurements.

**Conclusions**

Here we studied the growth behavior of *E. coli* at low enrichment levels of deuterium, and found that the adverse effects characteristic for high enrichment levels become noticeable at 0.5% D, which is several times higher level than expected given the 0.05% precision of our measurements. This discrepancy highlights the "buffering capacity" of living organisms that can partially compensate the adverse nature of the isotopically altered growth medium by reduced incorporation levels of heavy isotopes [28]. On the other hand, there seems to be a small but significant fingerprint of more comfortable environment at an ultralow enrichment level of 0.03% D. This behavior, which begs for additional verification by more precise measurements, could be explained by both hormesis as well as the isotopic resonance hypothesis. The latter theory predicts the exact location of the "resonance" at ≈0.03%-0.06% D, which is in a broad agreement with sporadic but multiple earlier reports [12–22].

**Acknowledgments**

The authors thank the Swedish Research Council for funding this work.

**FIGURES**

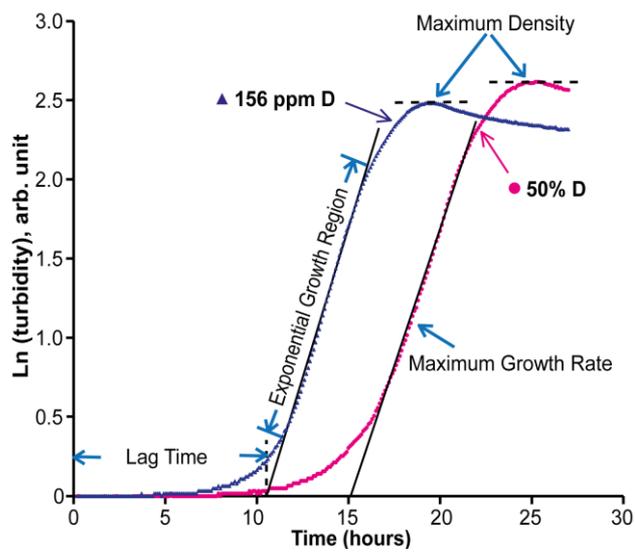

**Figure 1. Typical growth curve and the three growth parameters derived from the curve.**

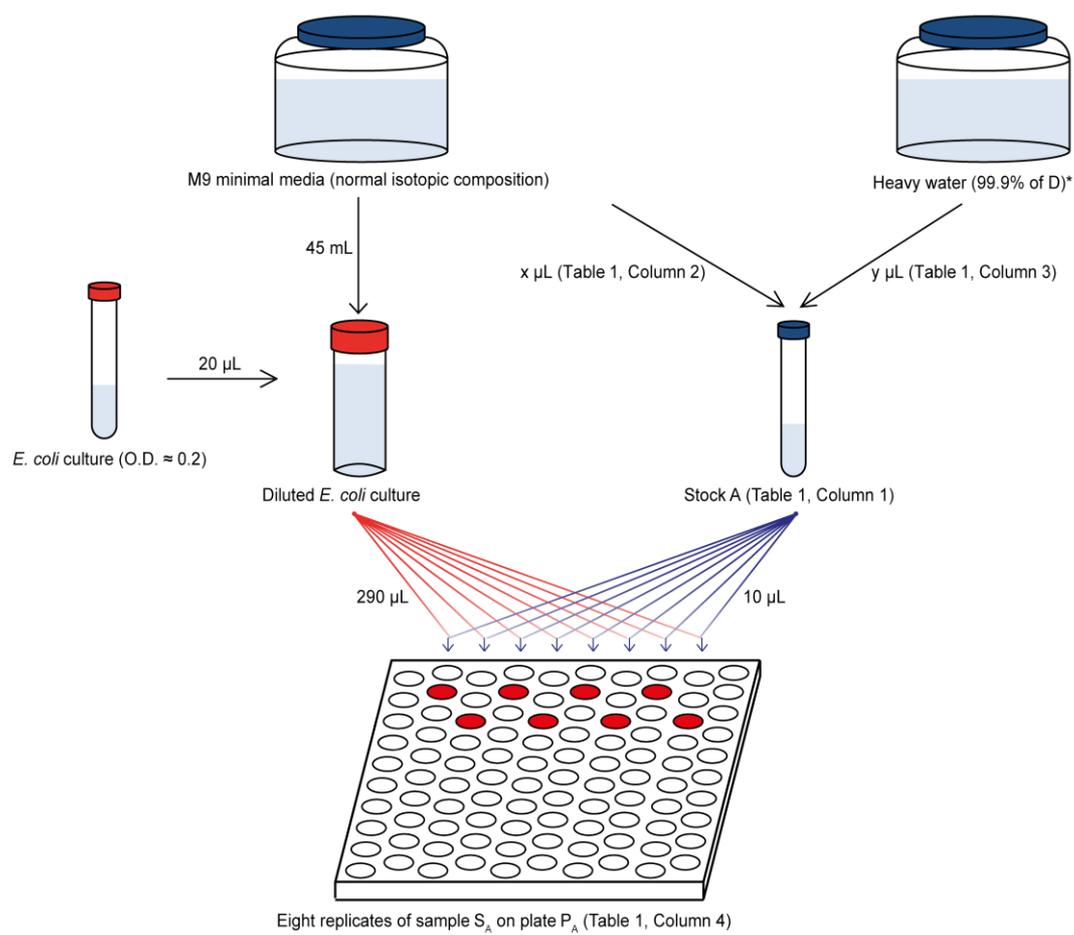

**Figure 2. Experimental workflow.** For each plate, 32 samples ($S_A$) and 32 standards were prepared. Stock A was used for preparing 32 samples on plate $P_A$.

* Milli-Q water was used instead of heavy water to prepare stock solution for the preparation of standards.

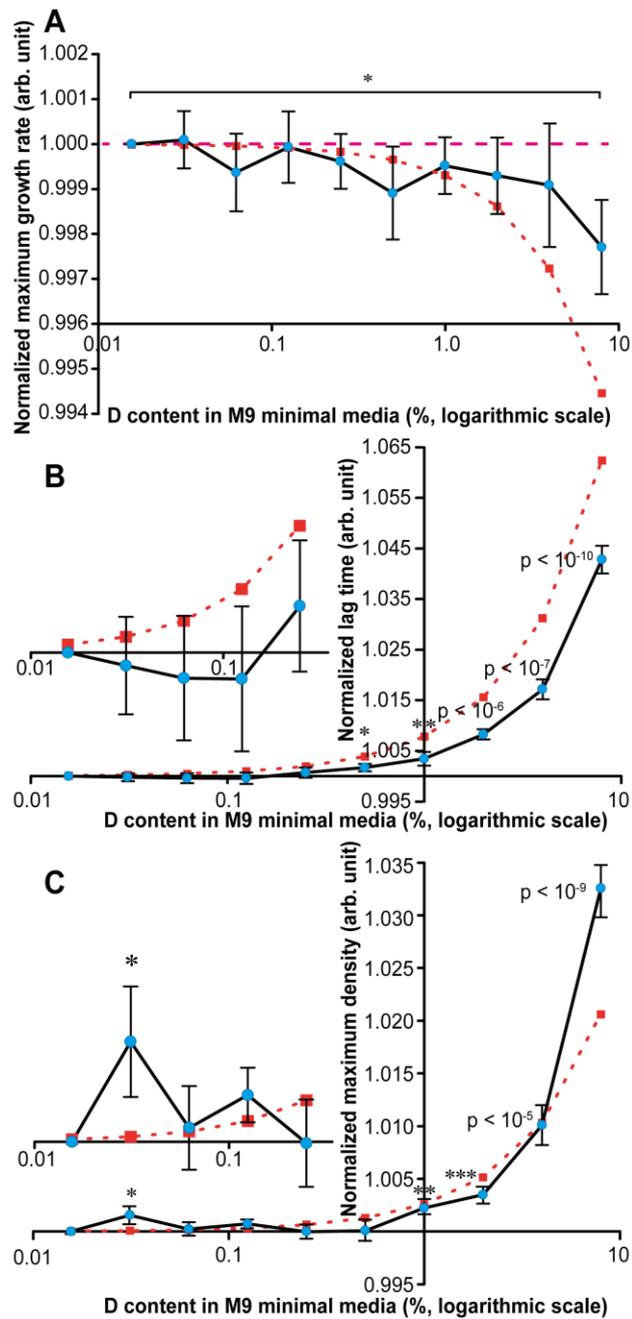

**Figure 3. Maximum growth rate, lag time, maximum density of *E. coli* grown in minimal media.** (**A**) Blue circles: maximum growth rate of *E. coli* grown in M9 minimal media normalized by that at normal deuterium content of 156 ppm. * denotes $p<0.05$, ** - $p<0.005$, etc. Brown squares: predicted maximum growth rate calculated according to the maximum growth rate of *E. coli* grown in 50% of deuterium. (**B**) Blue circles: lag time of *E. coli* grown in M9 minimal media normalized by that at terrestrial content of deuterium from 156 ppm (terrestrial value) to 8‰. Inset shows a zoom-in of the ultralow enrichment region. Brown squares: predicted lag time calculated according to the lag time of *E. coli* grown in 50% of deuterium. (**C**) Blue circles: maximum density of *E. coli* grown in M9 minimal media normalized by that at terrestrial deuterium content from 156 ppm (terrestrial value) to 8‰. Inset shows a zoom-in of the ultralow enrichment region. Brown squares: predicted maximum density calculated according to maximum density of *E. coli* grown in 50% of deuterium.

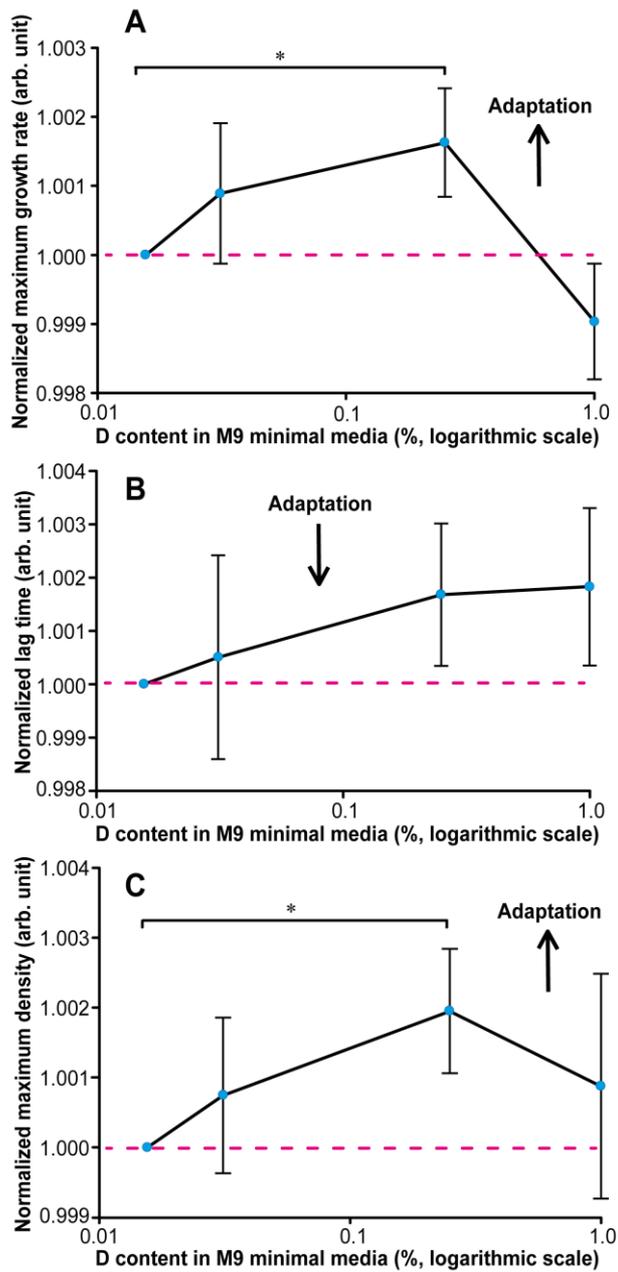

**Figure 4. Maximum growth rate, lag time and maximum density of aged *E. coli*.**
(**A**) Maximum growth rate of aged *E. coli* grown in M9 minimal media with ultralow composition of deuterium. * denotes p<0.05. (**B**) Lag time of aged *E. coli* grown in M9 minimal media with ultralow composition of deuterium. (**C**) Maximum density of aged *E. coli* grown in M9 minimal media with ultralow composition of deuterium. * denotes p<0.05.

**Table 1.** Stock solutions and their corresponding D content in the final samples.

| D content in stock solution | M9 minimal media (µL) | Heavy water (µL) | D content in the final sample (honey comb well plate) |
|---|---|---|---|
| 0.0156% | 5948 | 28 (Milli-Q water) | 0.0156% |
| 0.48% | 5948 | 28 | 0.03% |
| 1.42% | 5892 | 84 | 0.06% |
| 3.29% | 5957 | 202 | 0.12% |
| 7.04% | 5728 | 433 | 0.25% |
| 14.52% | 5620 | 955 | 0.5% |
| 29.50% | 4403 | 1844 | 1% |
| 59.46% | 2437 | 3581 | 2% |
| 39.80% | 3667 | 2427 | 4% |
| 79.73% | 1222 | 4830 | 8% |

Stock solution (column one) was prepared by mixing M9 minimal media and heavy water at a certain ratio that resulted in the final D content in the sample (column four). For each sample, stock solution for its corresponding standards (below it is called stock solution of standard X) was prepared in the same way but using Milli-Q water instead of heavy water.

**Supplementary Figures**

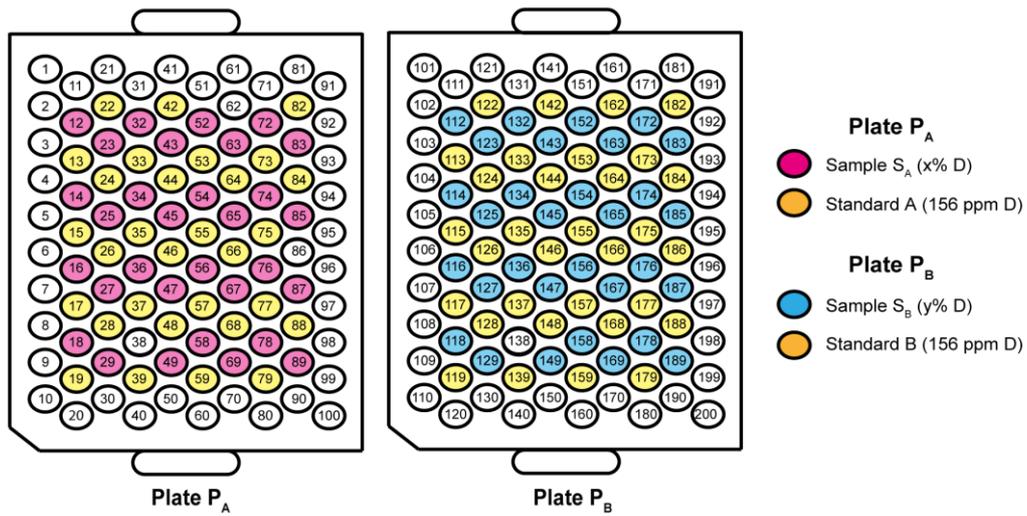

**Figure S1:** Sample configuration on the honey comb well plates.

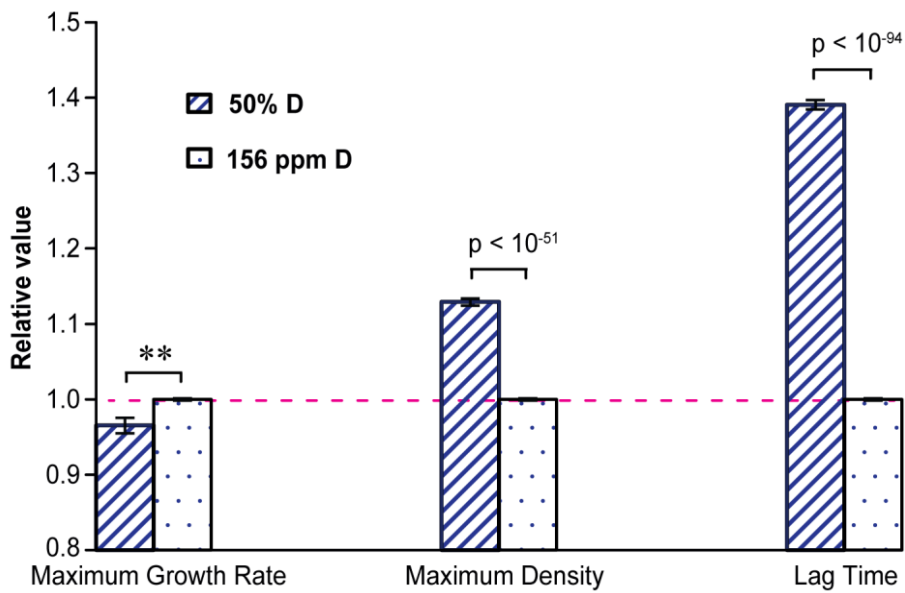

**Figure S2:** Maximum growth rate, maximum density and lag time of *E. coli* grown in M9 minimal media with deuterium content of 50% normalized by that at normal deuterium content of 156 ppm. ** is equivalent to p<0.005.

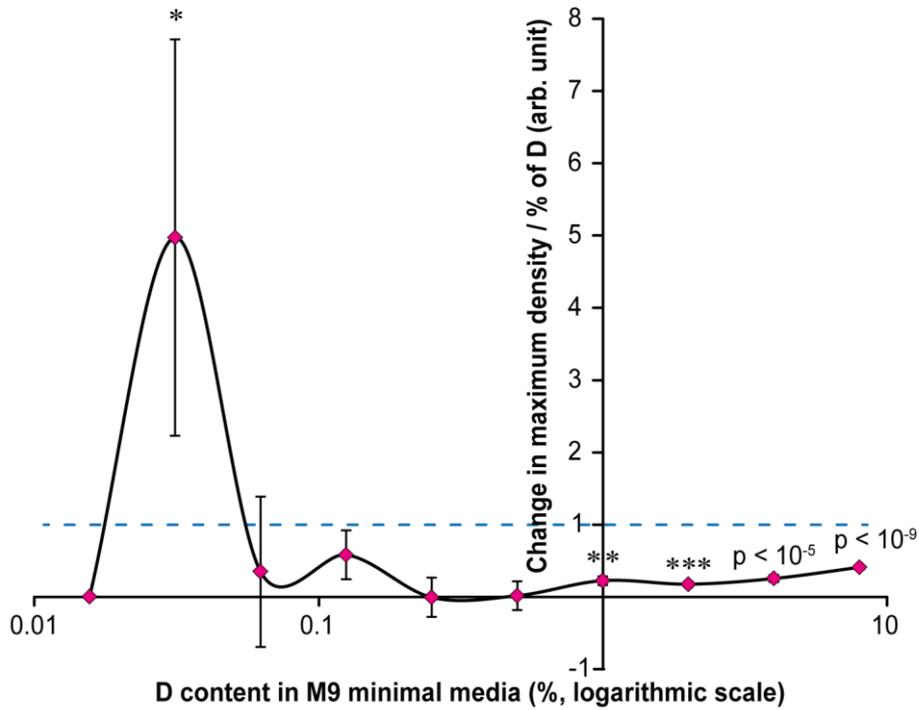

**Figure S3:** Change of maximum density per percentage of deuterium for *E. coli* grown in M9 minimal media with content of deuterium from 156 ppm (terrestrial value) to 8%. * is equivalent to p<0.05, ** is equivalent to p<0.005, etc.

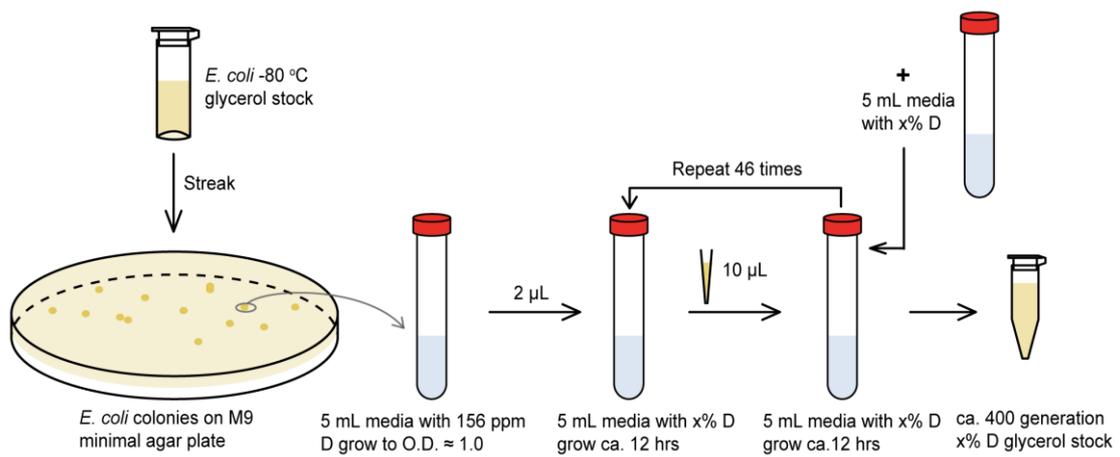

**Figure S4:** Workflow of adapting the bacteria to growth media.

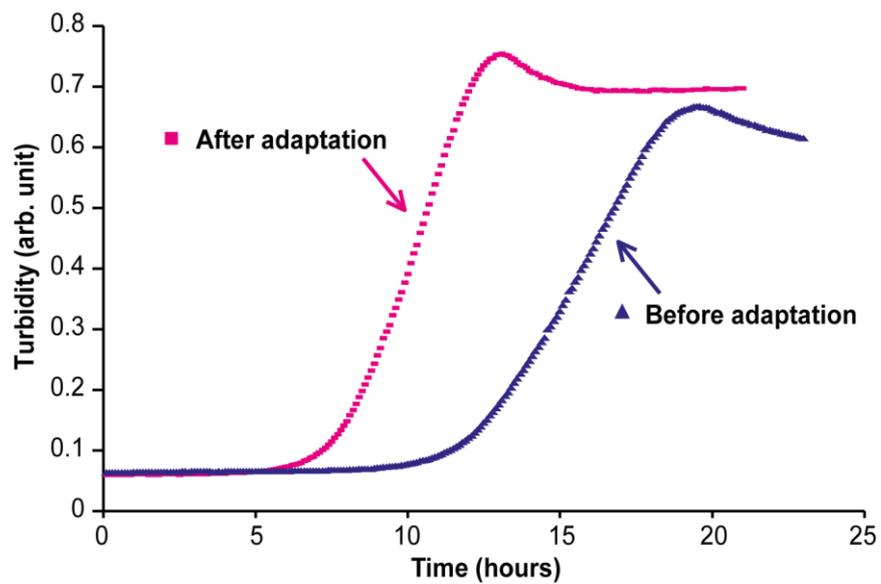

**Figure S5:** Growth curves of *E. coli* grown in minimal media with 156 ppm of D before and after adaptation**.**